  \documentclass[12pt]{article}
  \usepackage{times}

  \topmargin - 0.5 cm 
\begin{document}

  \baselineskip = 13 pt

\vspace*{3.5cm}

\begin{center}
{\large \bf A SU(2) recipe for mutually unbiased bases}  
\end{center}

\vspace{0.5cm}

\begin{center}
{\bf Maurice R.~Kibler}
\end{center}

\begin{center}
{Institut de Physique Nucl\'eaire de Lyon}\\
{IN2P3-CNRS/Universit\'e Claude Bernard Lyon 1}\\
{43 boulevard du 11 novembre 1918}\\
{F-69622 Villeurbanne Cedex, France}\\
{kibler@ipnl.in2p3.fr}
\end{center}

\begin{center}
and
\end{center}

\begin{center}
{\bf Michel Planat}
\end{center}

\begin{center}
{Institut FEMTO-ST, D\'epartement LPMO}\\
{CNRS/Universit\'e de Franche-Comt\'e}\\
{32 avenue de l'Observatoire}\\
{25044 Besan\c con Cedex, France}\\
{planat@lpmo.edu}
\end{center}

\newpage

\begin{center}
{\large \bf A SU(2) recipe for mutually unbiased bases}  
\end{center}

\vspace{0.5cm}

\begin{center}
{\bf Maurice R.~Kibler$^{a,}$\footnote{Corresponding author.

E-mail addresses: kibler@ipnl.in2p3.fr, planat@lpmo.edu}, Michel Planat$^b$}
\end{center}

\begin{center}
{$^a$Institut de Physique Nucl\'eaire de Lyon, 
IN2P3-CNRS/Universit\'e Claude Bernard Lyon~1, 
43 boulevard du 11 novembre 1918, F-69622 Villeurbanne Cedex, France}
\end{center}

\begin{center}
{$^b$Institut FEMTO-ST, D\'epartement LPMO, 
CNRS/Universit\'e de Franche-Comt\'e, 
32 avenue de l'Observatoire, F-25044 Besan\c con Cedex, France}

\end{center}

 \vspace{2.5cm}

  \noindent {\bf Abstract}

A simple recipe  for  generating  a complete set of mutually unbiased bases
in dimension $2j+1$,  with $2j$ integer and $2j + 1$ prime, 
is developed from a single matrix $V_a$ acting on a space of constant angular 
momentum $j$ and defined  in terms of the irreducible characters of the 
cyclic group $C_{2j+1}$. This recipe yields an (apparently new) compact formula 
for the vectors spanning the various mutually unbiased bases. In 
dimension $(2j+1)^e$,  with $2j$ integer, $2j + 1$ prime and $e$ positive integer,
the use of direct products of matrices of type $V_a$ makes it possible to
generate mutually unbiased bases. 
As two pending results, the matrix $V_a$ is used in
the derivation of a polar decomposition of SU(2) and of a FFZ algebra.

\vspace{0.5cm}



\newpage

\section{Introduction}

The notion of mutually unbiased bases (MUBs),$^{1-23}$ 
originally introduced by Schwinger 
and named by Wootters,$^{3}$ is of paramount 
importance in quantum information theory, especially in quantum 
cryptography and quantum state tomography. Let us recall that 
two orthonormal bases 
$\{ |a n_\alpha \rangle : n_\alpha = 0, 1, \cdots, d-1 \}$ and 
$\{ |b n_\beta  \rangle : n_\beta  = 0, 1, \cdots, d-1 \}$ 
of a $d$-dimensional Hilbert space, with an inner product denoted as 
$\langle \, | \, \rangle$, are said to be mutually unbiased if and only if 
$$
| \langle a n_\alpha | b n_\beta \rangle | = 
\delta(    a ,        b    )
\delta( n_\alpha , n_\beta ) + [1 - 
\delta(    a ,        b    )] \frac{1}{\sqrt{d}}.
$$ 
In dimension $d$, the maximum number of pairwise MUBs is $d+1$;$^{1-5}$ 
a set consisting of $d+1$ pairwise MUBs is called 
a complete set. As a matter of fact, 
the upper bound $d+1$ is attained when $d$ is a prime number 
or the power of a prime number.$^{2-9,16}$ There are numerous ways 
for constructing complete sets of MUBs,$^{1-23}$ most ot them 
being based on discrete Fourier analysis in Galois fields 
and Galois rings,$^{3,9,12,14,16,19,21}$ discrete Wigner functions,$^{3,10,21,22}$ 
generalized Pauli matrices.$^{5-8,10}$ Note 
also that the existence of MUBs can be 
related to the problem of finding mutually 
orthogonal Latin squares$^{11,15,22}$ and 
a solution of the mean King problem.$^{11,22}$ 
Let us also mention that the existence of MUBs has been
addressed by various authors from the point of view of 
finite geometries.$^{13,15,17,19}$ Finally, Lie algebra approaches 
to MUBs have been developed recently.$^{20,23}$

The main aim of this  note  is to give a simple algorithm for generating MUBs 
in dimension $d$ where $d$ is a prime number. The case where $d$ is the power 
of a prime number is briefly examined. The present work constitutes a
continuation of the ones in Ref.~23.

\section{The Main Results}

Let $\epsilon(j)$ be a $(2j+1)$-dimensional Hilbert space of constant angular momentum $j$
(the quantum number $j$ is such that $2j \in {\bf N}^*$). An orthonormal basis for 
$\epsilon(j)$ is provided by the set $\{ | j , m \rangle : m = j, j-1, \cdots, -j \}$ 
where the angular momentum state vectors $| j , m \rangle$, sometimes referred to as 
spherical or computational or Fock states, are eigenstates of the square $J^2$ 
of a generalized angular momentum and its $z$-component $j_z$. 

Following the suggestion made in Ref.~23 of ``redefining the operator $U_r$'' 
used in a study of SU(2), we introduce the 
$(2j+1)$-dimensional unitary matrix
\begin{eqnarray*}
V_a = 
\pmatrix{
0      &    q^a &      0  & \cdots &       0 \cr
0      &      0 & q^{2a}  & \cdots &       0 \cr
\vdots & \vdots & \vdots  & \cdots &  \vdots \cr
0      &      0 &      0  & \cdots & q^{2ja} \cr
1      &      0 &      0  & \cdots &       0 \cr
}, \quad a \in \{ 0, 1, \cdots, 2j \},
\end{eqnarray*}
builded on the spherical or standard basis 
$b_s = ( |j , j \rangle, |j , j-1 \rangle, \cdots, |j , -j \rangle )$. Here, 
the parameter $q$ is a rooth of unity defined by 
$$
q = \exp \left( {\rm i} \frac{2 \pi}{2j+1} \right).
$$
We have the immediate property
$$
{\rm Tr} \left( V_a ^{\dagger} V_b \right) = (2j+1) \delta (a,b).
$$
The matrix $V_a$ is a generalization of the matrix $U_r$ with $r \in {\bf R}$
considered in Ref.~24 in the 
framework of a polar decomposition of SU(2) and used 
in Ref.~23 for generating MUBs in the cases $d = 2$ and $3$. The set 
$\{ V_0, V_1, \cdots, V_{2j} \}$ of the $2j+1$ matrices $V_a$  
is constructed from the $2j+1$ irreducible character vectors of the cyclic group 
$C_{2j+1}$. Indeed, the nonzero matrix elements of the matrix $V_a$ are given by the 
irreducible character vector
$$
\chi^a = (1, q^a, \cdots, q^{2ja})
$$  
of $C_{2j+1}$.

It is straightforward to find the eigenvalues and eigenvectors of $V_a$. As a result, the 
spectrum of $V_a$ is non-degenerate. The eigenvector $|j a n_{\alpha} \rangle$
corresponding to the eigenvalue 
$$
\lambda (j a n_{\alpha}) = q^{ja - n_\alpha}
$$
reads
\begin{equation}
| j a n_{\alpha} \rangle = \frac{1}{\sqrt{2j+1}} \sum_{m = -j}^{j} 
q^{\frac{1}{2}(j + m)(j - m + 1)a + (j + m)n_\alpha} | j , m \rangle,
\end{equation}
where $n_{\alpha} = 0, 1, \cdots, 2j$. The $2j+1$ eigenvectors $| j a n_{\alpha} \rangle$ of the matrix $V_a$ 
generate an orthonormal basis $b_a$ of the space $\epsilon(j)$. For fixed $a$, 
the bases $b_a$ and $b_s$ are mutually unbiased. More specifically, we have 
the following result.

\vspace{\baselineskip}

\noindent {\bf Result 1}. In the case where $2j+1$ is a prime integer, the set 
comprizing the spherical basis $b_s$ and the $2j+1$ bases $b_a$ for 
$a = 0, 1, \cdots, 2j$ constitute a complete set of $2(j+1)$ MUBs.

\vspace{\baselineskip}

At this point, a natural question arises. How to construct a complete set of MUBs
for the direct product space 
$\epsilon(j) \otimes \epsilon(j) \otimes \cdots \otimes \epsilon(j)$ 
(with $e$ factors)
of dimension $d=(2j+1)^e$, where $2j+1$ is prime and $e$ is an 
integer greater or equal to 2? The answer 
follows from the following result.

\vspace{\baselineskip}

\noindent {\bf Result 2}. In the case where $2j+1$ is a prime integer, 
the eigenvectors of the matrices 
$$
W_{a_1 a_2 \cdots a_e} = V_{a_1} \otimes V_{a_2} \otimes \cdots \otimes V_{a_e}, \quad 
a_i \in \{ 0, 1, \cdots, 2j \}, \quad i = 1, 2, \cdots, e,
$$
together with the $d$-dimensional computational basis can be arranged to form 
a complete set of $d + 1 = (2j+1)^e + 1$ MUBs. 

\vspace{\baselineskip}

The proofs of Results 1 and 2 can be obtained 
from an adaptation of the proofs in Refs.~5-7, 12 and 21. The term ``arranged'' 
in Result 2 means that auxilliary matrices need to be introduced in order 
to deal with the degeneracy problem.  

As a corollary of Result 1, we otain the sum rule
$$
\left| \sum_{k = 0}^{d-1} 
q^{\frac{1}{2} k (d-k) (a-b) + k (n_\alpha - n_\beta)} \right| = 
d \delta(    a ,        b    )
  \delta( n_\alpha , n_\beta ) + \sqrt{d} [1 - \delta(    a ,        b    )],
$$
with
$$
q = \exp \left( {\rm i} \frac{2 \pi}{d} \right), 
\quad        a, b       \in \{ 0, 1, \cdots, d-1 \}, 
\quad n_\alpha, n_\beta \in \{ 0, 1, \cdots, d-1 \}, 
$$
where $d$ is a prime number.

\section{Two Related Results}

We would like to outline two Lie-like aspects of our approach. 

First, we 
can find a polar decomposition of the shift operators $j_+$ 
and $j_-$ of the Lie group SU(2) in terms of the unitary operator 
$v_a$ associated to the matrix $V_a$. The operator $v_a$ 
satisfies
$$
v_a | j , m \rangle = q^{(j-m)a} [1 - \delta(m,j)] |j , m+1 \rangle 
                      + \delta(m,j) | j , -j \rangle
$$
for $m = j, j-1, \cdots, -j$. Following Refs.~23 and 24, let us 
define the Hermitean operator $h$ through
$$
h | j , m \rangle = \sqrt{(j + m)(j - m + 1)} | j , m \rangle.
$$
We can show that the linear operators
$$
j_+ = h               v_a, \quad 
j_- = v_a^{\dagger} h,     \quad  
j_z = \frac{1}{2} ( h^2 - v_a^{\dagger} h^2 v_a )
$$
have the following action 
\begin{eqnarray}
j_{\pm} | j , m \rangle = 
q^{\pm(j \mp m + \frac{1}{2} \mp \frac{1}{2})a} \sqrt{(j-m)(j+m+1)} 
|j , m \pm 1\rangle, \quad  j_z | j , m \rangle = m | j , m \rangle
\end{eqnarray}
on the standard state vector $| j , m \rangle$ for $m = j, j-1, \cdots, -j$. As
a consequence, we get 
$$
[j_z , j_{\pm}] = \pm j_{\pm}, \quad [j_+ , j_-] = 2 j_z.
$$
Hence, the operators $j_+$, $j_-$ and $j_z$  
span the Lie algebra of SU(2). This result is to be compared 
with similar results obtained in Refs.~21 and 23-25 without the occurrence 
of the parameter $a$. It is to be emphasized that this result holds 
for any value of $a$ ($a = 0, 1, \cdots, 2j$).  However,  note that the action 
of $j_{\pm}$ on $| j , m \rangle$ depends on $a$. The Condon and Shortley phase 
convention used in atomic spectroscopy amounts to take $a = 0$ in Eq.~(2).

Second, the cyclic character of the irreducible representations 
of $C_{2j+1}$ renders possible to express $V_a$ in function of 
$V_0$. In fact, we have 
$$
V_a = V_0 Z^a,
$$
where 
$$
Z = {\rm diag} (1, q, \cdots, q^{2j}).
$$
The matrices $V_a$ and $Z$ have an interesting property, namely, 
they $q$-commute in the sense that 
$$
V_a Z - q Z V_a = 0.
$$
By defining
$$
T_m = q^{\frac{1}{2}m_1m_2} V_a^{m_1} Z^{m_2}, \quad m = (m_1 , m_2) \in {{\bf N}^*}^2,
$$
we easily obtain the commutator
$$
[T_m , T_n] = 2 {\rm i} \sin \left( \frac{\pi}{2j+1} m \wedge n \right) T_{m+n},
$$
where 
$$
m \wedge n = m_1 n_2 - m_2 n_1, \quad m + n = (m_1 + n_1 , m_2 + n_2),
$$
so that the linear operators $T_m$ span the FFZ infinite dimensional Lie 
algebra introduced by Fairlie, Fletcher and Zachos.$^{26}$ The latter result 
parallels the ones obtained, on one hand, from a study of $k$-fermions and 
of the Dirac quantum phase operator through a $q$-deformation of the harmonic 
oscillator$^{27}$ and, on the other hand, from an investigation 
of correlation measure for finite quantum systems.$^{25}$

\section{Closing Remarks}

In the case where $d=2j+1$ is a prime number, Result 1 
provides us with a simple mean for generating a complete 
set of $d+1$ MUBs from the knowledge of a single matrix, 
viz., the matrix $V_a$. It should be noted that when 
$2j+1$ is not a prime number, Eq.~(1) can be used for spanning 
MUBs as well; however, in that case, it is not possible to 
generate a complete set of MUBs. 

The main interest of our approach relies on the fact that 
MUBs can be constructed from a simple generic matrix $V_a$
and yields calculations easily codable on a computer. In 
addition, the matrix $V_a$ turns out to be of physical 
interest and plays an important role in the polar 
decomposition of SU(2) and for the derivation of the FFZ 
algebra. 

These matters, inherited from a $q$-deformation approach to
symmetry and super\-sym\-metry,$^{27,28}$ will be developed 
in a forthcoming paper in a larger context involving MUBs, 
useful in quantum information, and symmetry adapted bases, 
useful in molecular physics and quantum chemistry.

\section*{Acknowledgements}

One of the author (M.R.K.) is grateful to Prof. D. Ellinas for useful
correspondence.

\end{document}